Research article

# Application of single-cell deep learning in elucidating the mapping relationship between visceral and body surface inflammatory patterns

Haixiang Huang, Yeping Shi, Bingbing Shen, Zhenwei Zhang, Jianming Yue, Lu Mei, Qiusheng Chen *

*College of Veterinary Medicine, Nanjing Agricultural University, Nanjing, Jiangsu Province 210095, China*

ABSTRACT

The subcutaneous fascia (scFascia) is a widely distributed connective tissue, yet its contribution to systemic responses during visceral inflammation remains insufficiently defined. In the rat model of Salmonella-induced dysentery, scRNA-seq of scFascia combined with deep neural network (DNN)-based nonlinear feature discovery revealed transcriptional alterations indicative of Warburg-like reprogramming in macrophages (MPs), accompanied by reduced activity of the citrate cycle. Cd34 + /Pdgfra+ telocytes (CPTCs) regulated MP differentiation and proliferation through Wnt/Fgf signaling, suggesting a pathological communication pattern between scFascia and visceral organs, termed the fascia–visceral inflammatory crosstalk pattern (FVICP). PySCENIC analysis showed increased activity of transcription factors Fosl1, Nfkb2, and Atf4, modulated by CPTC signaling to MPs, leading to downregulation of aerobic respiration and upregulation of the cell cycle, DNA replication, and transcription. Co-culture assays demonstrated that CPTCs suppressed the pro-inflammatory activity of MPs by transcriptionally downregulating TLR4 and CPT1A while upregulating ATF4, and enhanced anti-inflammatory potential via post-transcriptional MTOR upregulation. These findings highlight the role of scFascia in immunomodulation and metabolic reprogramming during visceral inflammation, emphasizing its contribution to systemic homeostasis.

## 1. Introduction

Inflammatory lesions release chemokines that mediate myeloid cell trafficking from the bone marrow and spleen to affected tissues through trans-endothelial recruitment [1]. However, the complex mechanisms governing immune cell migration remain poorly understood. Traditional phenotypic studies have primarily focused on localized changes within pathological tissues, often neglecting the broader implications for systemic homeostasis.

The scFascia, a pervasively distributed loose connective tissue enriched with mesenchymal and immune cells, has been extensively studied in wound healing and angiogenesis [2,3]. Due to its intricate vascular, neural, and cellular networks, scFascia is uniquely suited to receive and integrate signals from visceral pathologies [4,5]. However, its function in maintaining systemic balance, particularly in inflammatory conditions, has not been fully elucidated.

Among the diverse interstitial cell subpopulations within fascia, CPTCs have emerged as a distinctive population with notable morphological characteristics [6]. Formally defined by Popescu and Faussone-Pellegrini in 2010[7], their identification rests on three criteria: (1) immunophenotypic markers showing CD34 + /PDGFRα+ double-positive expression; (2) morphology characterized by long cell bodies with cytoplasmic projections; and (3) ultrastructural features revealing nanometer-scale thread-like cytoplasmic extensions under transmission electron microscopy. These criteria have since been validated across organs including the heart, lung, gastrointestinal tract, and urogenital system [8]. CPTCs play critical roles in tissue repair [9], intercellular communication [10], and organ homeostasis maintenance [11]. They establish extensive communication networks with surrounding cells through direct contact, extracellular vesicle release, and secretion of soluble factors.

Salmonella enteritis, an acute gastroenteritis caused by invasive Salmonella infection, exemplifies a pathological condition involving the interplay between intestinal microbiota and cells within the digestive tract [12]. Upon pathogen invasion, intestinal macrophages undergo Warburg-like metabolic reprogramming—glycolysis under aerobic

* Corresponding authors.
*E-mail addresses:* alex-hxhuang@stu.njau.edu.cn (H. Huang), chenqsh305@njau.edu.cn (Q. Chen).

conditions—marked by reduced aerobic metabolism and disruption of the TCA cycle. At sites of dysenteric lesions, Salmonella breaches host defenses when the intestinal flora is disrupted, invades macrophages, hijacks metabolic regulatory pathways during intracellular proliferation, and redirects host metabolism toward pathogen-favorable conditions, thereby inducing Warburg-like reprogramming [13]. This mechanism requires distinction from the Warburg-like reprogramming observed in LPS-induced M1 polarization of pathogen-free macrophages [14]. Warburg reprogramming reflects a shift in cellular energy metabolism [15], commonly driven by heightened demands for DNA synthesis during cellular proliferation [16], immune evasion [17], or, in tumor cells, dedifferentiation [18]. The emergence of Warburg-like reprogramming thus signals enhanced proliferative capacity and complex immunological remodeling [12]. This phenomenon is particularly relevant in Salmonella enteritis, where acute inflammation strongly modulates immune cell proliferation and mobilization to counteract pathogenic threats [19].

In this study, a Salmonella-induced enteritis model was established in Sprague–Dawley (SD) rats to investigate the systemic influence of visceral inflammation on scFascia. We characterized transcriptional alterations in scFascia cell populations in response to visceral disease using scRNA-seq integrated with DNN-based nonlinear feature discovery. This approach uncovered the previously unrecognized homeostatic regulatory role of scFascia and its functional significance in systemic inflammatory pathologies.

## 2. Methods and materials

### 2.1. Ethical approval

The experimental subjects, female Sprague-Dawley rats aged eight weeks and weighing approximately 250 g, were obtained from Beijing Vital River Laboratory Animal Technology Co., Ltd. Ethical clearance for all animal-related experimental protocols was secured from the Institutional Animal Care and Use Committee of Nanjing Agricultural University, under protocol number NJAU.No20220427087. This study strictly adhered to the Institutional Animal Care and Use Committee's guiding principles and aligned with the Animal Research: Reporting *In Vivo* Experiments guidelines. The rodents were housed in a fully ventilated cage system, ensuring unrestricted access to food and water. Environmental conditions were stringently maintained with a stable ambient temperature of approximately 25°C and a natural light-dark cycle to simulate the animals' native habitat.

### 2.2. Bacterial strains and growth conditions

The Salmonella Typhimurium strain CVCC542 (hereafter referred to as S. Typhimurium CVCC542) was obtained from the Joint International Research Laboratory of Animal Health and Food Safety. The cultivation process was conducted in Luria-Bertani (LB) broth (Beyotime, # ST156), supplemented with 50 μg/mL kanamycin, at 37°C. Initial culturing involved overnight incubation in a 2-mL static culture at 37°C to select the motile bacteria. Subsequently, the upper 1 mL of this culture was transferred into 49 mL of preheated LB broth and subjected to shaking incubation for 5 h at the same temperature. The optical density of the S. Typhimurium culture was assessed at 600 nm using an Eppendorf BioPhotometer Plus to ascertain the bacterial concentration. An appropriate bacterial dose was then prepared in sterile 0.01 M phosphate-buffered saline (PBS) with a pH of approximately 7.4.

### 2.3. Animal assays

Six female Sprague-Dawley (SD) rats, aged eight weeks with an average body weight of 250 g, were subjected to a 3-day oral gavage regimen to induce Salmonella Typhimurium infection. The bacterial suspension was adjusted to a concentration of 5 × 10ˆ8 colony-forming units per milliliter using 0.01 M PBS. To neutralize gastric acidity prior to infection, the dysentery group (DG) received an oral dose of 5 % sodium bicarbonate ($NaHCO_3$). Ten minutes later, 400 μL of the bacterial suspension was administered via gavage. This procedure was repeated every 24 h for three consecutive days at 2:00 PM. The healthy group (HG) consisting of an additional six SD rats, was maintained on a standard diet without intervention.

### 2.4. Sampling

This section delineates the methodological framework for preparing tissue samples of DG and HG, facilitating morphological analysis and single-cell sequencing investigations.

Procedures:

Sacrifice and Compliance:

All subjects were euthanized in accordance with the Institutional Animal Care and Use Committee protocols, and under Animal Research: Reporting *In Vivo* Experiments guidelines to uphold ethical standards.

Morphological Studies:

Specimens, comprising intact skin and scFascia along the abdominal midline with a width of 2 cm from the pubic symphysis to the xiphoid, were harvested for paraffin sections and transmission electron microscopy analysis.

Fixation: Samples underwent fixation in a solution containing 4 % paraformaldehyde and 0.25 % glutaraldehyde in PBS, ensuring the preservation of cellular architecture for subsequent detailed morphological assessment.

Preparation for Single-cell Sequencing:

The scFascia was sampled along the abdominal midline with a width of 2 cm from the pubic symphysis to the xiphoid, meticulously dissected to isolate it from surrounding dermal and adipose tissues.

Storage: To preserve cellular viability before tissue dissociation, specimens were maintained in a 0°C ice bath using a specialized tissue storage solution (Miltenyi Biotec, # 130–100–008).

### 2.5. Immunofluorescence, Hematoxylin, and Eosin staining

The hydrated paraffin-embedded tissue sections underwent antigen retrieval through incubation in 0.01 M sodium citrate at 98°C for 5 min. Post-incubation, sections were treated with primary antibodies, followed by fluorescently labeled secondary antibodies and DAPI staining (Beyotime, # C1005). Fluorescent images were acquired using a confocal microscope (Carl Zeiss, LSM900). Subsequently, sections were dehydrated for hematoxylin and eosinstaining and preserved with neutral resin for extended storage.

Antibodies employed in this investigation:

Anti-CD34 (Abclonal, # A13929)

Anti-PDGFRα (Santa Cruz, # sc-398206)

Anti-F4/80 (Abcam, # Ab300421)

Alexa Fluor 488 Anti-Rabbit Fc Fragment (Proteintech, # SA00013–2)

Alexa Fluor 594 Anti-Mouse Fc Fragment (Proteintech, # SA00013–3)

### 2.6. Transmission electron microscopy

Fascia samples were initially fixed in a 2.5 % glutaraldehyde solution for 48 h at 4°C, followed by post-fixation in 1 % osmium tetroxide for 2 h under the same temperature conditions. The specimens underwent dehydration through a graded alcohol series, comprising concentrations of 75 %, 85 %, 95 %, and 100 %, each for 10 min. Subsequently, the tissues were embedded in epoxy resin. Ultrathin sections, approximately 50 nm thick, were mounted on copper grids and subjected to staining with uranyl acetate and lead citrate. The prepared sections were then analyzed using a Hitachi HT7800 transmission electron microscope operating at an acceleration voltage of 80 kV.

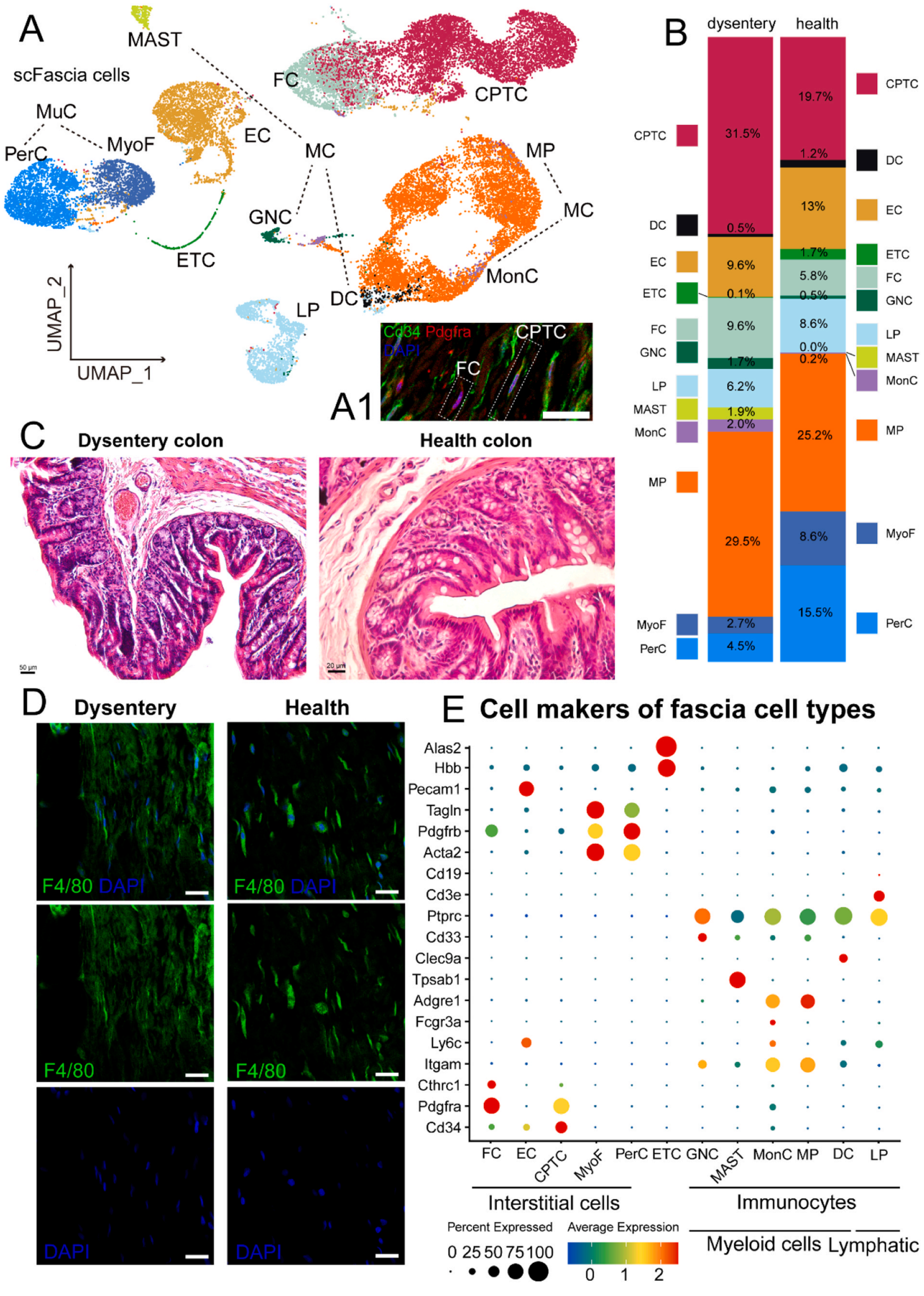

(caption on next page)

**Fig. 1.** UMAP visualization of scRNA-Seq data from subcutaneous fascial tissues in DG and HG.A. UMAP projection of scRNA-Seq data from subcutaneous fascial tissues of rats with Salmonella-induced dysentery and healthy controls, illustrating the separation of cellular clusters based on disease status : Broad categories: Cd34 + /Pdgfra+ Telocytes (CPTCs), Myeloid cells (MCs), Fibroblasts (FCs), Endothelial cells (ECs), Mural cells (MuCs), Lymphatic cells (LPs), Hbb+ erythroid cells (ETCs); Fine subpopulations: MC: Monocytes (MonCs), Macrophages (MPs), Dendritic cells (DCs), Neutrophils (NEUs), Mast cells (MAST); MuC: Pericytes (PerCs), Myofibroblasts (MyoFs)A1. Validation for cell markers of CPTCs and FCs with immunofluorescence assays in scFascia. Double immunofluorescence staining shows CD34 + (green) and PDGFRα+ (red) double-positive CPTCs distributed in parallel with FC cells (CD34-/PDGFRα+) within the fascial matrix (scale bar: 20 µm).B. Percentage of cells expressing each gene, indicating the prevalence of specific markers in different cell populations between dysentery and healthy conditions. Numbers in the figure represent the cellular proportion of corresponding cell types within each sample group.C. Hematoxylin, and Eosin -stained colonic pathological alterations in Salmonella dysentery. The dysentery group shows obvious vascular congestion in the lamina propria and inflammatory infiltration in the submucosa, while the healthy group appears relatively clean.D. Macrophages (MPs) in the scFascia of dysentery and healthy groups. F4/80 + macrophages exhibit distinct morphological differences between groups. Macrophages in dysentery tissues display elongated, spindle-shaped morphology with extensive cytoplasmic processes, while healthy controls show rounded, compact cellular morphology with minimal projections (scale bar: 20 µm).E. Average expression levels of the highlighted genes in the identified cell populations, highlighting differential expression associated with cell types.

### 2.7. Tissue dissociation and cell purification for scRNA-seq

Initially, the tissue was transferred from the Miltenyi storage solution into a sterile culture dish containing 10 mL of 1x Dulbecco's Phosphate-Buffered Saline (DPBS; Thermo Fisher, # 14190144) on ice. This step was implemented to eliminate residual blood and adipose tissue, after which the tissue was finely minced on ice.

A specialized dissociation solution was prepared, comprising collagenase type 4 (Sigma-Aldrich, # C5138), elastase (Sigma-Aldrich, # E1250), and 10 µg/mL DNase I (Thermo Fisher, # EN0523), all dissolved in PBS supplemented with 5 % Fetal Bovine Serum (FBS; Thermo Fisher, # SV30087.02). The tissue was subjected to enzymatic digestion in this solution at 37°C with gentle agitation at 50 rpm for approximately 40 min. Dissociated cells were collected at 20-minute intervals to optimize cell yield and viability.

The resultant cell suspension underwent sequential filtration through nylon cell strainers decreasing from pore sizes of 100 µm to 40 µm to remove debris. The remaining red blood cells were lysed using 1X Red Blood Cell Lysis Solution (Thermo Fisher, # 00–4333–57). The dissociated cells were washed with 1x DPBS containing 2 % FBS. Cell viability was assessed using 0.4 % Trypan blue staining and confirmed with a Countess® II Automated Cell Counter (Thermo Fisher). Subsequently, cell suspensions were processed with the Dead Cell Removal Kit (Miltenyi Biotec, # 130–090–101), ensuring that only samples with greater than 95 % viability were advanced for further analysis.

To construct 10X libraries and 3' sequencing, suspension cells were prepared using a water-in-oil solution. This solution was crafted by the Chromium™ Single Cell Controller (10X Genomics), enabling high-fidelity capture and sequencing of individual cells.

10X library construction

In this study, beads, each imbued with a unique molecular identifier (UMI) and cell barcodes, were loaded to near saturation, ensuring each cell was paired with a bead within a Gel Beads-in-emulsion (GEM). Upon exposure to a cell lysis buffer, polyadenylated RNA molecules hybridized to these beads and were subsequently collected into a single tube for reverse transcription. During cDNA synthesis, each cDNA molecule was uniquely tagged at the 5′ end, corresponding to the 3′ end of an mRNA transcript, with a UMI and a cell-specific label. The beads underwent sequential processes of second-strand cDNA synthesis, adaptor ligation, and universal amplification. Sequencing libraries were prepared from randomly fragmented whole-transcriptome amplification products to enrich the 3′ ends of transcripts associated with cell barcodes and UMIs, adhering to the standard protocol of the Chromium Single Cell 3′ v3.1 system. The libraries were precisely quantified using both a High Sensitivity DNA Chip (Agilent) via a Bioanalyzer 2100 and the Qubit High Sensitivity DNA Assay (Thermo Fisher Scientific), culminating in sequencing on an Illumina NovaSeq 6000 platform using $2 \times 150$ bp read chemistry.

### 2.8. scRNA-seq data preprocessing and quality control

The expression matrix was generated by Cellranger 7.2.0. Single-cell RNA sequencing data from DG and HG fascial tissues were processed using customized R scripts using Seurat v4.4 within the R environment version 4.3.0. Raw 10X Genomics data were imported to create Seurat objects. Quality control was performed with sample-specific thresholds:

1. DG samples: Cells with $< 28{,}804$ UMIs, $< 6260$ genes, and $< 12$ % mitochondrial content were retained; 2. HG samples: Cells with $< 31{,}018$ UMIs, $< 7201$ genes, and $< 10$ % mitochondrial content were retained. All samples required a minimum expression of 50 genes per cell and detection in $\geq 3$ cells per gene. Automated outlier removal and pre/post-QC visualization were implemented before generating final quality-controlled Seurat objects.

### 2.9. Single-cell Data Integration and Clustering

Quality-controlled Seurat objects from DG and HG fascial tissues were integrated using the standard Seurat workflow. Datasets were merged with condition labels (HG vs DG), normalized via LogNormalize, and subjected to dimensionality reduction (PCA followed by UMAP/tSNE). Graph-based clustering was performed at multiple resolutions (0.2–1.5), with cluster-specific marker genes identified using Wilcoxon rank-sum tests.

Pre-defined marker gene panels were applied in cell type annotation. Cell identities were annotated through a two-tiered system:

1. **Broad categories** (celltype): Cd34 + /Pdgfra+ Telocytes(CPTCs), Myeloid cells (MCs), Fibroblasts (FCs), Endothelial cells (ECs), Mural cells (MuCs), Lymphatic cells (LPs), Hbb+ erythroid cells (ETCs);
2. **Fine subpopulations** (celltype2): MC: Monocytes (MonCs), Macrophages (MPs), Dendritic cells (DCs), Neutrophils (NEUs), Mast cells (MAST); MuC: Pericytes (PerCs), Myofibroblasts (MyoFs)

Differential gene expression (DGE) was analyzed using the following framework:

1. Sample-level: Health vs dysentery comparisons across all cell types
2. Subpopulation-level: Condition-specific differences within celltype2 categories (MonC, MP, PerC, etc.)

Statistical filtering retained genes with $|\log_2 FC = > 0.25$ and adjusted p-value $< 0.05$.

### 2.10. Deep learning model architecture and training strategy

#### 2.10.1. Model architecture

We adopted a Multi-Layer Perceptron (MLP) architecture for single-cell expression-based nonlinear feature discovery tasks. The model structure includes:

1. Input Layer: High-dimensional gene expression features (input dimension equals the number of genes); 2. Hidden Layers: 4–7 fully connected layers, with 64–128 nodes per layer (standard search) or 20–59 nodes (fine-tuned search); 3. Output Layer: Number of nodes equals the number of cell types, no explicit activation function

**Table1**
Representative marker genes expression in major cell types.

| Broad Categories | Fine Categories | Gene | log2FC | p_val_adj |
|---|---|---|---|---|
| **CPTC** | CPTC | Pdgfra | 1.21 | < 1.00E−300 |
| | | Cd34 | 0.81 | < 1.00E−300 |
| **FC** | FC | Pdgfra | 1.72 | < 1.00E−300 |
| | | Cthrc1 | 0.67 | < 1.00E−300 |
| **EC** | EC | Pecam1 | 3.01 | < 1.00E−300 |
| **MC** | MP | Cd68 | 2.54 | < 1.00E−300 |
| | | Cd14 | 2.51 | < 1.00E−300 |
| | MonC | Fcgr3a | 1.67 | 1.09E−282 |
| | | Cd14 | 1.10 | 1.05E−64 |
| | MAST | Tpsab1 | 5.49 | < 1.00E−300 |
| | DC | Cd74 | 1.90 | 2.75E−103 |
| | | Clec9a | 1.03 | < 1.00E−300 |
| **MuC** | MyoF | Tagln | 4.24 | < 1.00E−300 |
| | | Acta2 | 3.30 | < 1.00E−300 |
| | PerC | Acta2 | 2.48 | < 1.00E−300 |
| | | Pdgfrb | 1.95 | < 1.00E−300 |
| **LP** | LP | Cd3d | 1.86 | < 1.00E−300 |
| **ETC** | ETC | Hbb | 11.01 | 9.10E−295 |
| | | Alas2 | 4.05 | < 1.00E−300 |

Notes:
CPTC: Cd34 + /Pdgfra+ telocytes; FC: Fibroblasts; EC: Endothelial cells; MC: Myeloid cells; MuC: Mural cells; GNC: Granulocytes; LP: Lymphocytes; ETC: Hbb+ Erythroid cells
MP: Macrophages; MonC: Monocytes; MAST: Mast cells; MyoF: Myofibroblasts; PerC: Pericytes; DC: Dendritic cells
log2FC: average $\log_2$ fold change; p_val_adj: adjusted p-value

(probability calculation integrated in loss function);

#### *2.10.2. Hyperparameter optimization*

We conducted a comprehensive grid search across the following hyperparameter space:

*2.10.2.1. Standard hyperparameter search space.* 1. Learning Rate: 0.01–0.09 (step size: 0.01); 2. Hidden Layer Dimensions: 64, 128 nodes; 3. Network Depth: 4–7 hidden layers; 4. Optimizers: SGD, Adam, RMSprop; 5. Hidden Layer Activation Functions: ReLU, Sigmoid, ELU, Softmax, Hardswish; 6. Batch Size: 32 (fixed); 7. Loss Function: Cross-entropy loss (with built-in softmax computation); 8. Total Parameter Combinations: 1080 combinations.

*2.10.2.2. Fine-tuned hyperparameter search space.* 1. Learning Rate: 0.01–0.09 (step size: 0.01); 2. Hidden Layer Dimensions: 20–59 nodes (continuous integer range); 3. Network Depth: 4–7 hidden layers; 4. Optimizers: SGD, Adam, RMSprop; 5. Hidden Layer Activation Functions: ReLU, Sigmoid, ELU, Softmax, Hardswish; 6. Batch Size: 32 (fixed); 7. Loss Function: Cross-entropy loss (with built-in softmax computation); 8. Total Parameter Combinations: 21,600 combinations.

#### *2.10.3. Overfitting prevention strategies*

To ensure model generalization capability, we implemented multiple regularization techniques:

1. Early Stopping Mechanism: Training terminates if validation accuracy does not improve for 10 consecutive epochs; 2. Learning Rate Scheduling: StepLR scheduler with step size of 30 epochs and decay factor of 0.1; 3. Data Splitting: 90 % training and 10 % independent testing to ensure unbiased performance evaluation; 4. Cross-validation: Independent batch validation (b1 and b2) for robust performance assessment.

#### *2.10.4. Optimal model parameters*

Through systematic hyperparameter optimization, we identified cell-type-specific optimal configurations:

| Cell Type | Learning Rate | Hidden Layer Nodes | Hidden Layers | Optimizer | Activation |
|---|---|---|---|---|---|
| CPTC | 0.01 | 128 | 5 | Adam | ELU |
| MC | 0.01 | 128 | 4 | Adam | ELU |
| EC | 0.01 | 31 | 7 | RMSprop | Hardswish |
| ETC | 0.01 | 44 | 5 | RMSprop | ELU |
| FC | 0.01 | 128 | 4 | Adam | ReLU |
| MuC | 0.01 | 128 | 4 | Adam | Hardswish |
| LP | 0.01 | 128 | 4 | Adam | ELU |

Independent whole-fascia cross-validation batches (b1 and b2) from different independent biological replications achieved the following optimal parameters:

| Batch | Learning Rate | Hidden Layer Nodes | Hidden Layers | Optimizer | Activation Function |
|---|---|---|---|---|---|
| b1 | 0.01 | 64 | 4 | Adam | ELU |
| b2 | 0.03 | 64 | 5 | SGD | ELU |

#### *2.10.5. Model interpretability*

The feature importance was assessed through DeepSHAP (Deep SHapley Additive exPlanations), which calculates each gene’s contribution to classification decisions. SHAP values were computed for all genes across all cells, then aggregated using mean absolute values across samples and classes to generate gene importance rankings. All genes entering downstream Gene Set Enrichment Analysis (GSEA) must simultaneously satisfy two conditions: 1. SHAP score $> 0$ (indicating positive contribution to feature learning in the DNN model); 2. Significant in Differential Gene Expression (DGE) analysis (showing significant expression changes before and after dysentery).

### *2.11. Group and pipeline design of deep learning and model interpretation*

We annotated single-cell sequencing data from scFascia, comparing DG and HG. The expression matrices for each cell population were normalized, scaled, and labeled for nonlinear feature discovery tasks. During the training phase, a primary class of MLP was defined at first, and an extensive parameter search for batch size, hidden layers, activation functions, and optimizers was conducted to identify the optimal model parameters based on validation dataset performance. These parameters were then used to generate models for subsequent interpretation and visualization. Model interpretation employed the DeepSHAP [20] method, ranking feature genes by their contribution to model accuracy. Genes with SHAP values were filtered and used to refine differentially expressed gene (DEG) lists, which were subsequently subjected to Gene Set Enrichment Analysis (GSEA) to identify biologically relevant functions uncovered by deep learning.

In cross-validation experiments using whole fascia single-cell data, biological replicates were isolated, and the expression matrices for the entire scFascia were processed identically to generate models. To assess generalization performance, models trained on one biological replicate were tested on datasets from other replicates, serving as unseen data. The results were visualized using confusion matrices to evaluate accuracy.

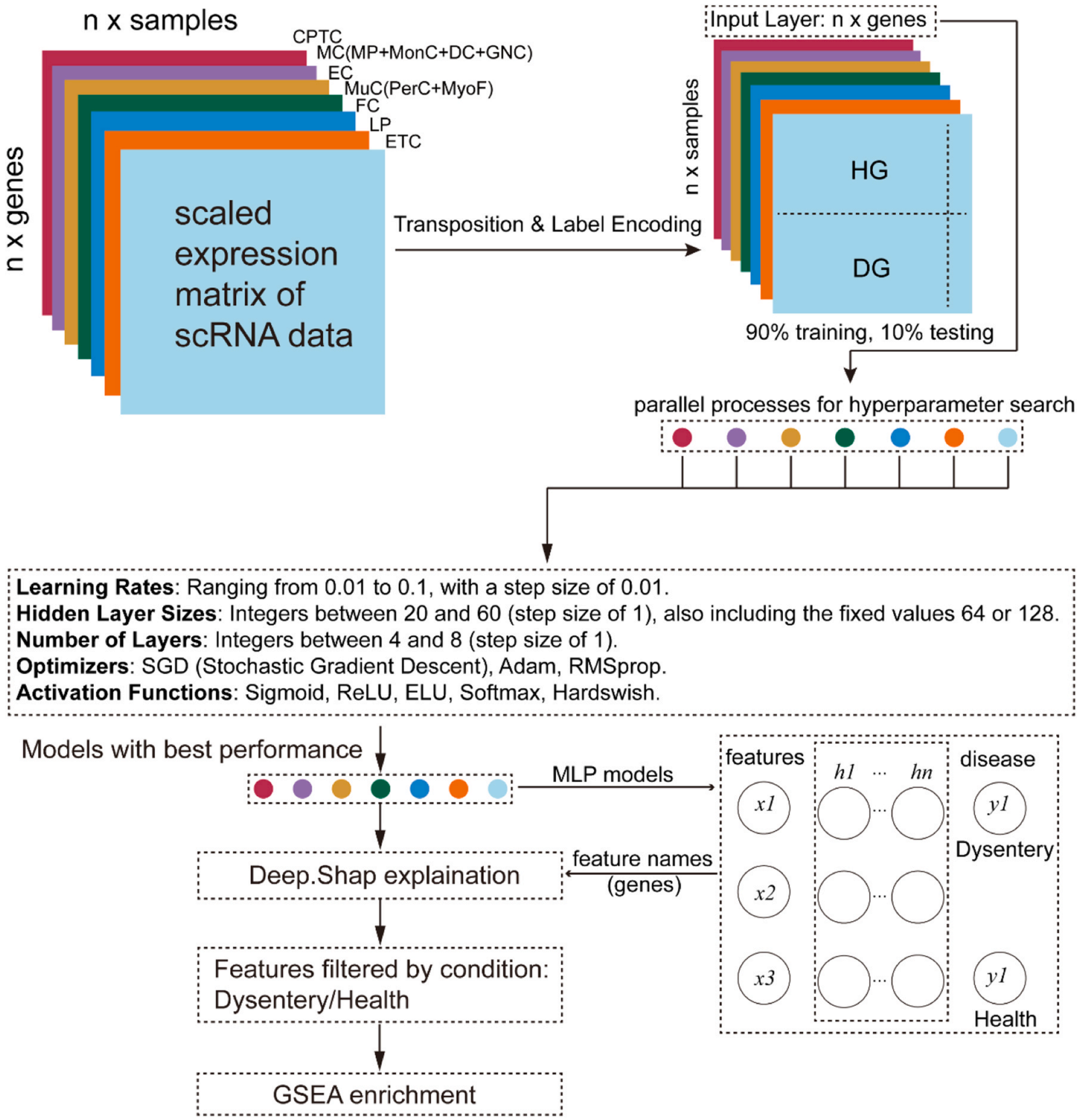

**Fig. 2.** Schematic overview of the deep learning pipeline implemented in this study.The pipeline begins with the scaled expression matrix of single-cell RNA (scRNA) data (n genes x n samples), representing gene expression profiles across various cell populations: CPTC, MC(MP+MonC+DC+GNC), EC, MuC(PerC+MyoF), FC, LP, ETC. This matrix undergoes transposition and label encoding, transforming it into an input layer of n genes x n samples, which is then split into 90 % training HG and 10 % testing DG sets.A parallel hyperparameter search is conducted to optimize the Multi-Layer Perceptron (MLP) models. The search space for hyperparameters includes:1. Learning Rates: Ranging from 0.01 to 0.1, with a step size of 0.01.2. Hidden Layer Sizes: Integers between 20 and 60 (step size of 1), and fixed values of 64 or 128.3. Number of Layers: Integers between 4 and 8 (step size of 1).4. Optimizers: Stochastic Gradient Descent (SGD), Adam, and RMSprop.5. Activation Functions: Sigmoid, Rectified Linear Unit (ReLU), Exponential Linear Unit (ELU), Softmax, and Hardswish.Overfitting Prevention Strategies:1. Early stopping: Training stops if validation accuracy does not improve for 10 consecutive epochs.2. Learning rate decay: Learning rate decays every 30 epochs (gamma=0.1).3. Train/test split: 90 % training, 10 % testing to ensure generalization.The models with best performance from the hyperparameter search are then subjected to DeepSHAP (Deep SHapley Additive exPlanations) explanation to determine the importance of individual features (genes) in the model's predictions. The models with best performance from the hyperparameter search are then subjected to DeepSHAP (Deep SHapley Additive exPlanations) explanation to determine the importance of individual features (genes) in the model's predictions. These important features are subsequently integrated with differential gene expression (DGE) results and filtered using dual criteria: genes must exhibit both positive SHAP values (>0, indicating ML-model importance) and statistical significance in traditional differential expression analysis ($p_val < 0.05$) to ensure biological relevance. This dual-criteria approach identifies genes that are both computationally informative for disease classification and biologically significant in traditional statistical testing, yielding robust nonlinear features specifically relevant to the disease state. Finally, Gene Set Enrichment Analysis (GSEA) is performed on the filtered features to identify enriched biological pathways and functions.

### 2.12. *Single-cell metabolic pathway activity analysis*

Single-cell metabolic inference is based on the scMetabolism package (https://github.com/wu-yc/scMetabolism), which infers cellular metabolic activity states by analyzing the expression patterns of metabolism-related genes in single-cell RNA sequencing data. This method utilizes the KEGG metabolic pathway database combined with the VISION algorithm to calculate metabolic pathway activity scores.

### 2.13. *PySCENIC regulon network inferring*

scFascia and colon count matrix was generated from the Seurat

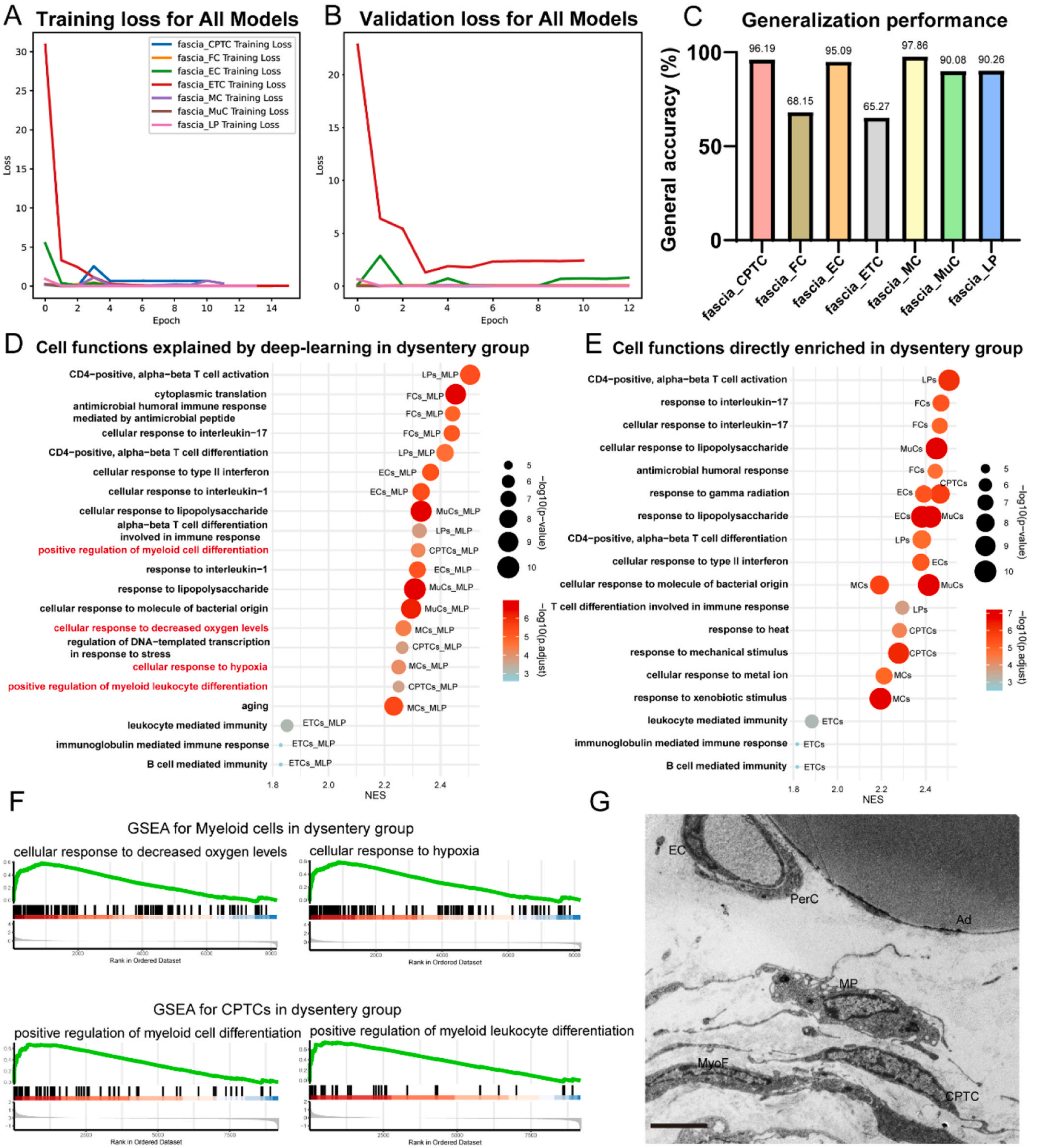

**Fig. 3.** Cell functions were extracted through deep learning based on neural network models.A. Training epochs and loss values for each cell population during deep learning. All cell types except ETCs showed relatively stable feature recognition capabilities.B. Validation loss values for each cell population across training epochs. C. Generalization performance of the optimal models for each cell population when applied to the full scFascia dataset. Notably, MCs, CPTCs, and ECs all achieved accuracy rates above 95 %, while FCs, despite having relatively abundant sample numbers, showed lower accuracy rates along with ETCs, both falling below 70 %.D. Top Gene Set Enrichment Analysis (GSEA) terms derived from filtered gene symbols extracted using the DeepSHAP algorithm, following the calculation of intergroup expression differences.E. Top GSEA terms obtained from directly computed DEGs for each cell population.F. GSEA enrichment distribution for myeloid cells (MCs) and CPTCs.G. Cellular composition of the perivascular microenvironment in scFascia under dysentery conditions. The spatial relationship between CPTCs with elongated cell body processes and macrophages with abundant pseudopod structures can be observed to be closely associated. Additionally, the major cell types in scFascia all appear within this narrow transmission electron microscopy field of view, confirming the very close *in* Vivo distribution of these cellular components (scale bar: 4μm).

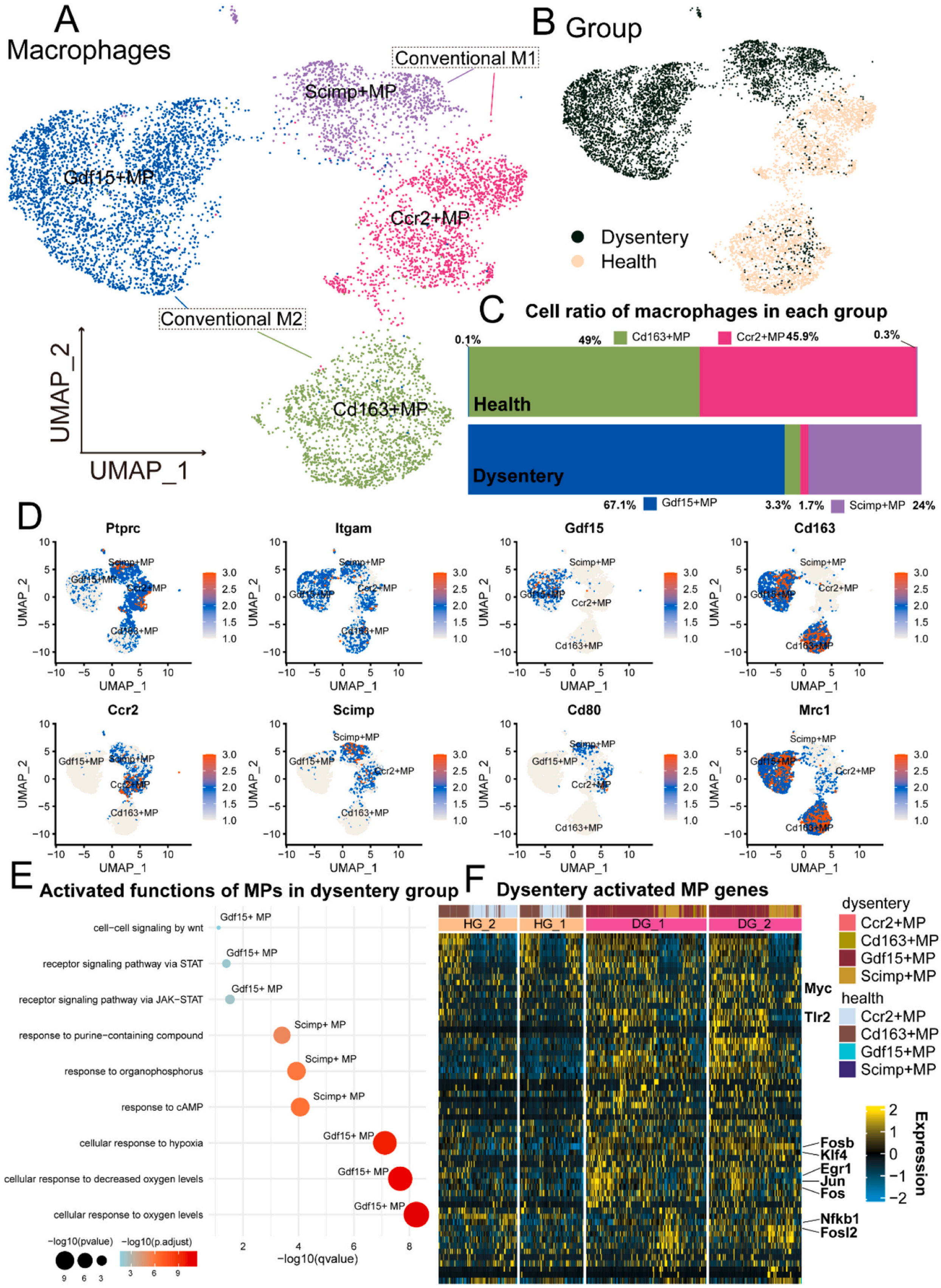

(caption on next page)

**Fig. 4.** Macrophage subpopulations and their functional characteristics.A. Annotation of macrophage populations (MPs) based on characteristic markers. Macrophages conforming to traditional M2 subtypes showed Gdf15 + and Cd163 + marker expression in DG and HG respectively, while those conforming to traditional M1 subtypes exhibited Scimp+ and Ccr2 + marker expression in DG and HG respectively.B. MP populations annotated according to group-specific differences. A portion of M2 macrophages from DG can still be observed within the Cd163 + macrophage cluster, while HG cells are barely visible in the Gdf15 + macrophage cluster.C. Proportion of marker-specific macrophage subsets within each group. Numbers represent the specific cellular proportions of these cell types in the corresponding groups.D. Expression of classical immune markers and group-specific genes in MPs, colors in the figure represent the scaled gene expression levels of each cell.E. DG-specific upregulated cellular functions in MPs exhibiting classical M2 characteristics.F. Heatmap showing the expression of genes associated with the functions highlighted in panel E across the entire macrophage population. Colors in the figure represent the scaled gene expression levels of each cell.

object after quality control as the input dataset, all procedures are guided by the GitHub project of PySCENIC (https://github.com/aertslab/pySCENIC). During the analysis pipeline, PySCENIC automatically performs motif-based pruning and implicit significance screening. All output TF regulons indicate they have passed the filtering based on default thresholds.

### 2.14. Cell communication analysis

We applied the quality-controlled Seurat objects of scFascia and colon from each group to generate Cellcall objects following the guide of this GitHub project (https://github.com/ShellyCoder/cellcall).

### 2.15. CytoTRACE analysis

To predict the relative differentiation states of macrophage populations, we employed CytoTRACE (Cellular Trajectory Reconstruction Analysis using gene Counts and Expression), a computational method that leverages the inverse relationship between the number of detectably expressed genes per cell and cellular differentiation status [21]. CytoTRACE analysis was performed using the CytoTRACE R package (version 1.0) on raw count matrices from macrophage populations.

### 2.16. THP-1 derived M2 macrophage culture

The THP-1 monocyte cell line was seeded in 12-well plates at $1 \times 10^5$ cells/mL using RPMI-1640 medium (Gibco, #11875093) supplemented with 10 % FBS. Differentiation into M0 macrophages was induced by 48-hour treatment with 100 nM PMA (MCE, #HY-18739). For M2 polarization, macrophages were stimulated with 20 ng/mL interleukin-4 (Novoprotein, #CK15) for 48 h to establish adherent M2 macrophages. Before the Transwell co-culture, the medium was replaced with a 2:1 mixture of RPMI-1640 and Ham's F-12 nutrient mixture (containing 10 % FBS) for 12-hour incubation.

### 2.17. Primary CPTCs extraction and culture

Complete subcutaneous fascial tissue specimens were collected according to anatomical principles consistent with morphological experiments. Tissue spanning from the xiphoid cartilage to the pubic symphysis was collected as 2-cm-wide symmetrical strips centered along the abdominal midline. After dissecting and mincing the scFascia, tissues were digested at 37°C with an enzymatic solution containing collagenase type 4 (Solarbio, # C8160), Dnase (Solarbio, # D8071), and hyaluronidase (Solarbio, # H8030). and elastase (Solarbio, # E8210). Sequential filtration through 100-μm and 40-μm meshes was performed, followed by red blood cell lysis solution (Biosharp, # BL503A) treatment. Suspension cells from the SF were cultured in DMEM/F12 medium (Gibco, # 11330032) supplemented with 10 % Fetal Bovine Serum (FBS; Gibco, Cat. No. 10099–141 C) and 1 % penicillin-streptomycin (Solarbio, Cat. No. P7630). The cells were incubated in a cell incubator (Thermo Fisher) at 37 ℃, 95 % humidity, and 5 % carbon dioxide. Differential plating was conducted at 2 h, 2–4 h, and > 4 h intervals, with cells adhering during the 2–4 h interval selected as the CPTC for co-culture studies. For Transwell assays, CPTCs were trypsinized and seeded in 0.4 μm Transwell insert chambers (Corning, # 3460) using RPMI-1640/F12 (2:1) medium containing 10 % FBS. After 12 h of incubation, co-culture experiments were initiated.

### 2.18. CPTC-M2 transwell co-culture assay

Transwell inserts seeded with CPTC were emptied of medium and transferred to 12-well plates containing M2 macrophages in the bottom chambers. Wells were replenished with fresh basal medium (0.5 mL in inserts, 1 mL in lower chambers). Co-culture was performed using RPMI-1640/F-12 nutrient mixture (2:1 ratio, 10 % FBS). Inflammation was induced by supplementing the mixture medium with 1 μg/mL lipopolysaccharide (LPS; Sigma-Aldrich, #L2880). After 24-hour co-culture, samples were harvested for M2 macrophage phenotyping.

### 2.19. Immunocytochemistry assay

Primary CPTCs isolated by differential time-based adhesion were fixed after 96-hour adherence. Medium was aspirated, and cells on coverslips were fixed with 4 % paraformaldehyde (PFA) at 4°C for 30 min. After PBS washing, specimens were sequentially treated with:

1. 1 % Triton X-100 (10 min incubation, RT)
2. Primary antibody (12 h incubation, 4°C)
3. Secondary antibody (1 h incubation, RT)
4. DAPI (10 min incubation, 37°C)

All steps were interspersed with two washes using 0.01 M PBS (5 min, 100 rpm orbital shaking). Coverslips were mounted in antifade mounting medium (Beyotime, # P0126) and fluorescence imaging was performed using an Olympus DP73 microscope equipped with a 40 × objective.

Antibodies employed in this investigation:
Anti-CD34 (Abclonal, # A13929)
Anti-PDGFRα (Santa Cruz, # sc-398206)
Anti-WNT2(Abclonal, # A5864)
Anti-CD34 (Santa Cruz, # sc-74499)
Alexa Fluor 488 Anti-Rabbit Fc Fragment (Proteintech, # SA00013–2)
Alexa Fluor 594 Anti-Mouse Fc Fragment (Proteintech, # SA00013–3)

### 2.20. RNA extraction and cDNA preparation

RNA was extracted from PBS-washed cell samples using TRIzol reagent (Vazymes #R401–01). Lysates underwent phase separation by adding chloroform (0.3 mL per 1 mL lysate), vortexing vigorously for 15 sec, and incubating at RT for 3 min. Centrifugation at 12,000 × g (10 min, 4°C) separated aqueous/organic phases.

The aqueous phase was transferred to RNase-free tubes, mixed 1:1 with ice-cold isopropanol, and incubated at 4°C for 30 min. RNA pellets were collected by centrifugation (12,000 × g, 10 min, 4°C), washed twice with 75 % ethanol (1 mL, centrifugation at 12,000 × g for 5 min at 4°C), and air-dried for 8 min. Pellets were resuspended in 40 μL DEPC-treated water and stored at −80°C.

RNA integrity was verified via NanoDrop™ (Thermo Fisher) with acceptance criteria:

- A260/A280: 1.8–2.0

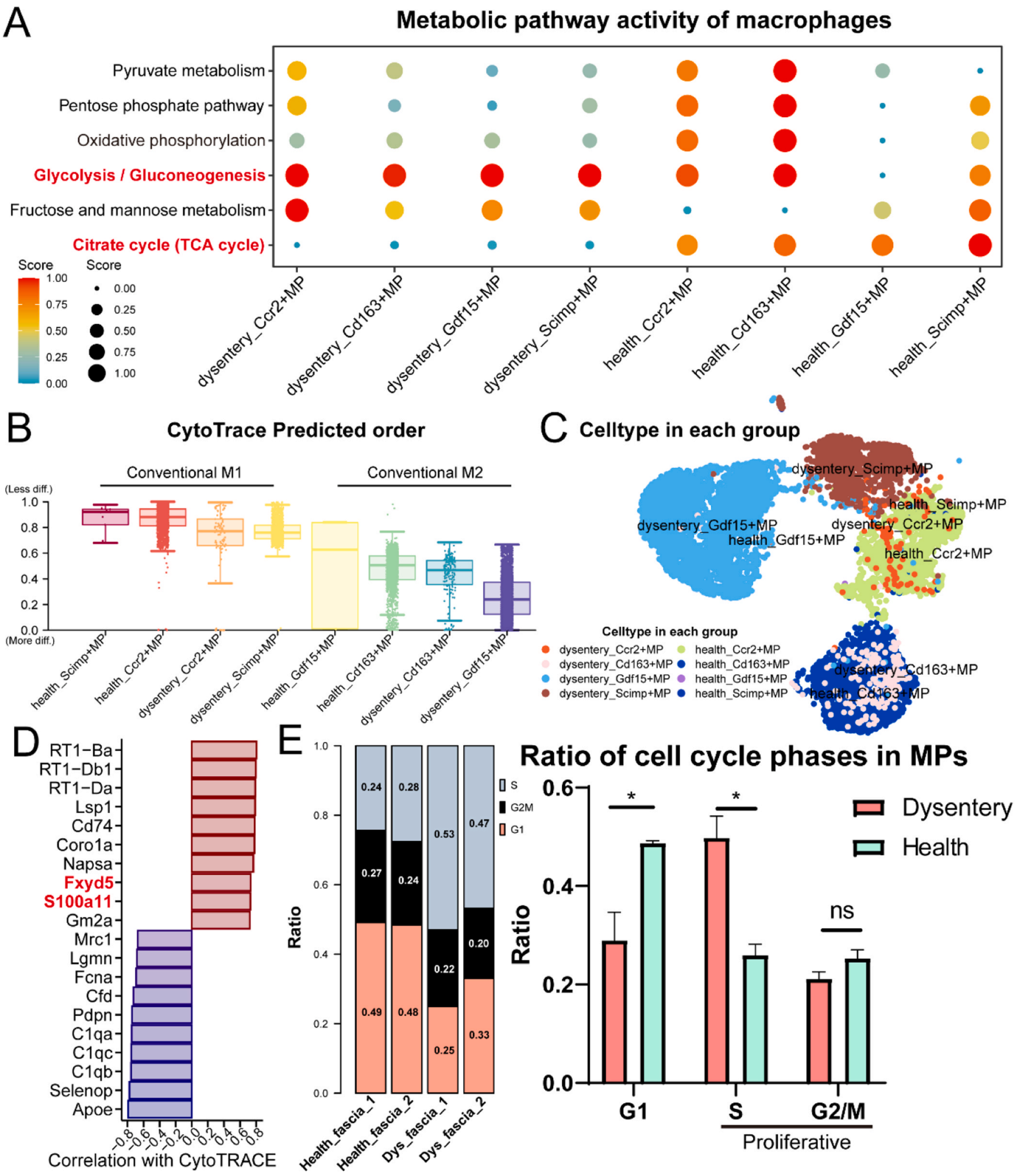

(caption on next page)

**Fig. 5.** Single-cell metabolic analysis and proliferative-differentiation characteristics of MPs.A. Metabolic pathway activity scores related to cellular respiration in each macrophage population (MP) subset across groups, as determined by scMetabolic analysis. Both color intensity and dot size represent the overall activity score of each metabolic pathway, with larger and redder dots indicating higher pathway activity (score range: 0.00–1.00). Key pathways highlighted include Glycolysis/ Gluconeogenesis and Citrate cycle (TCA cycle) showing distinct group-specific patterns.B. CytoTRACE scoring of MPs, where higher scores indicate lower differentiation states (more stem-like characteristics) and lower scores indicate higher differentiation states (more mature characteristics). Box plots show the distribution of CytoTRACE scores across different MP subtypes, with conventional M1 subtypes (left, warmer colors) generally showing higher scores than conventional M2 subtypes (right, cooler colors).C. UMAP visualization of MPs, with cells colored and annotated by both experimental group (dysentery vs health) and subset identity. This two-dimensional representation reveals the spatial distribution and clustering patterns of different MP subtypes, with clear separation between group-specific populations.D. Top 10 genes showing the highest positive correlation (red bars, upper panel) and strongest negative correlation (blue bars, lower panel) with CytoTRACE scores. Bar length represents the correlation coefficient strength, with longer bars indicating stronger correlations. Genes like S100a11, Fxyd5 show positive correlation (associated with stemness), while genes like Selenop, Apoe show negative correlation (associated with differentiation).E. Proportion of cell cycle phases within MPs, reflecting their proliferative activity. Bar charts show the relative proportions of G1 (pink), S (blue), and G2/M (black) phases in dysentery vs healthy groups. Statistical significance is indicated by asterisks (*), with dysentery MPs showing significantly altered cell cycle distribution (Fighser's test, $p < 0.05$), particularly increased S phase activity indicative of enhanced proliferation.

- A260/A230: 2.0–2.5

Reverse transcription was used with HiScript® II Reverse Transcriptase (Vazymes #R212–01), following the manufacturer's protocol to generate cDNA for qPCR. All $H_2O$ reagents used in this experiment were DEPC-pretreated (Beyotime, # R0021)

### 2.21. Real-time quantitative PCR (qPCR)

qPCR was performed strictly following the manufacturer's protocol using ChamQ Universal SYBR qPCR Master Mix (Vazymes, #Q711–02) on a Bio-Rad CFX96 system. Statistical analysis was conducted with Bio-Rad CFX Maestro software and GraphPad Prism 9. All primer sequences are listed below:

| Target | Primer sequence |
|---|---|
| CPT1A-F | CATGTCCAGCCAGACGAAGA |
| CPT1A-R | TGCTCAGTGAACATCCACCC |
| TFAM-F | AGTGGCGGGCATGATAACAC |
| TFAM-R | CCACATGCTTCGGAGAAACG |
| IL6-F | AGTGAGGAACAAGCCAGAGC |
| IL6-R | AGCTGCGCAGAATGAGATGA |
| TLR4-F | TACAGAAGCTGGTGGCTGTG |
| TLR4-R | ACCCGCAAGTCTGTGCAATA |
| ATF4-F | TCAGTCCCTCCAACAACAGC |
| ATF4-R | TCTGGCATGGTTTCCAGGTC |
| HPRT1-F | CCTGGCGTCGTGATTAGTGA |
| HPRT1-R | CGAGCAAGACGTTCAGTCCT |

### 2.22. Western blot (WB)

After washing with PBS, cell pellets were iced and lysed in RIPA buffer (Thermo Fisher, #89900) with gentle agitation at 15-minute intervals for 30 min. Lysates were mixed with SDS-PAGE loading buffer (Beyotime Biotechnology, #P0015L) containing 1 mM PMSF (Beyotime, #ST506) and DNase I (Solarbio, #D8071), denatured at 98°C for 10 min, and loaded onto 12.5 % polyacrylamide gels (Epizyme Biomedical Technology, #PG113). Electrophoresis proceeded at 80 V for 30 min followed by 120 V for 60 min. Proteins were wet-transferred to PVDF membranes (Labselect, #TM-PVDF-R-45) at 0.3 A constant current for 100 min. Membranes were blocked in 5 % skim milk for 1 h at RT, probed with primary antibodies overnight at 4°C, and subsequently incubated with HRP-conjugated secondary antibodies for 1 h at RT. After PBS washing, protein bands were detected using ECL substrate (Epizyme, #SQ201) and imaged.

Antibodies employed in this investigation:

Anti-MTOR (Abclonal, # A11355)

Anti-HPRT1 (Abclonal, # A21825)

Goat Anti-Rabbit Fc Fragment HRP-conjugated (ZSGB-BIO, # ZB-2301)

Goat Anti-Mouse Fc Fragment HRP-conjugated (ZSGB-BIO, # ZB-2305)

## 3. Results

### 3.1. Cellular proportion and inflammatory phenotype alterations in subcutaneous fascia under dysentery conditions

An expression matrix was generated using Cell Ranger version 7.2.0. A total of 30,388 high-quality single-cell profiles were obtained from the scFascia of the healthy group (HG: HG_1 and HG_2, n = 13,224) and the dysentery group (DG: DG_1 and DG_2, n = 17,164). No batch correction was required because the principal component analysis (PCA) reduction correlated with cell type annotations (Supplementary Figure 1 A, B).

Both interstitial and immune cells were annotated across cohorts, with all cell types distinctly distributed in UMAP coordinates, reflecting heterogeneous cellular states (Fig. 1A). The cellular populations within fascia were classified into 12 distinct types (Fig. 1A, B, E; Table 1).

To validate the presence of macrophages and CPTCs, immunofluorescence staining for Cd34/Pdgfra and F4/80 was performed in scFascia, revealing a spatially parallel distribution pattern of these cell types. CPTCs exhibited elongated morphology with slender cytoplasmic projections, with Cd34 localized to the cell surface and co-localized with Pdgfra. These cells were clearly distinguished from adjacent Pdgfra+ / Cd34- fibroblasts (FC), which displayed different localization patterns (Fig. 1A1). F4/80 + macrophages displayed distinct morphologies between groups: in dysentery tissues, they were elongated and spindle-shaped with extensive cytoplasmic processes, whereas in healthy controls they were rounded and compact with minimal projections (Fig. 1D). Histopathological examination of the colon, the primary target tissue, showed vascular congestion and infiltration after Salmonella gavage, although the mucosal epithelium remained largely intact (Fig. 1C).

For an overarching assessment, all annotated cells were grouped into six major categories: myeloid cells (MCs), CPTCs, endothelial cells (ECs), fibroblasts (FCs), lymphocytes (LPs), and Hbb⁺ erythroid cells (ETCs) (Fig. 1A; Table 1). Differential gene expression (DGE) analysis was conducted between dysentery and healthy groups for each cell type, followed by Gene Set Enrichment Analysis (GSEA). Results indicated that these cell types collectively exhibited inflammatory transcriptional signatures and pathogen-associated molecular pattern (PAMP)-driven innate immune responses. Comparison of the top three GSEA GO terms ranked by normalized enrichment scores (NES) revealed broad overlap in inflammatory characteristics, with some functional redundancy across cell types (Fig. 3E). For example, both mural cells (MuCs) and ECs showed activation in response to lipopolysaccharide (LPS), while MCs and MuCs exhibited responses to bacterial-derived molecules. These results confirmed that dysentery systemically elevated circulating PAMP concentrations, such as LPS [22].

Given the cellular composition of the perivascular microenvironment in scFascia (Fig. 3G), it is hypothesized that pathological macrophage-associated microenvironments contribute to homeostatic disruption, consistent with findings in prior studies [23]. These observations suggest that scFascia may participate in distal immune regulation during

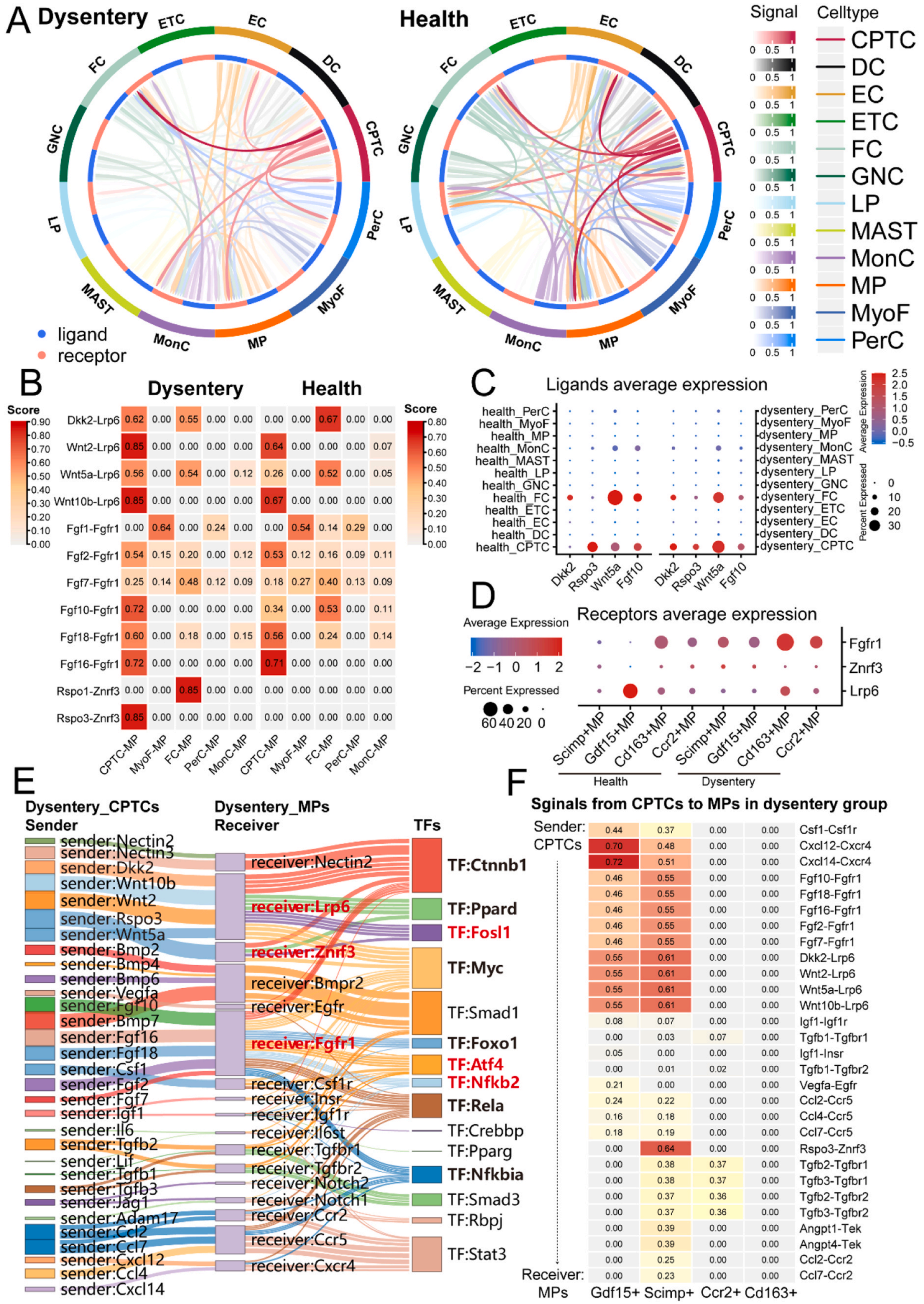

(caption on next page)

**Fig. 6.** Cell-cell communication analysis between MPs and CPTCs.A. Global visualization of cell-cell communication in scFascia from both groups using CellCall analysis. Arrow color intensity represents the communication probability score (ranging from 0 to 1), which quantifies the likelihood of ligand-receptor interaction based on co-expression patterns between sender and receiver cells.B. Heatmap dissecting Wnt/Fgf signaling pathways, highlighting differences in intercellular communication between dysentery and healthy groups. Color intensity represents the CellCall signaling strength score, calculated from ligand-receptor expression levels and downstream pathway activity; darker red colors indicate stronger predicted downstream transcriptional effects (score range: 0–1).C. Expression levels of ligand genes associated with the Wnt/Fgf interactions shown in panel B. Dot size represents the percentage of cells expressing each ligand, while color intensity indicates the average normalized expression level (log-transformed and scaled).D. Expression levels of receptor genes associated with the Wnt/Fgf interactions shown in panel B. Dot size represents the percentage of cells expressing each receptor, while color intensity indicates the average normalized expression level (log-transformed and scaled).E. Sankey diagram illustrating the regulatory cascade of cell-cell communication between CPTCs (senders) and MPs (receivers). From left to right: ligand expression in CPTCs, corresponding receptors in MPs, and predicted downstream transcription factor (TF) activities. Line thickness represents the strength of predicted regulatory relationships based on CellCall communication scores.F. Heatmap of CPTCs-MPs communication signals stratified by macrophage subpopulations. Color intensity represents the CellCall communication score for each ligand-receptor pair in specific MP subtypes, with red indicating stronger predicted downstream transcriptional activation (score range: 0–1). Numbers in cells show the exact communication strength values.

**Table 2**
Cell functions enriched by positively expressed Gdf15 + MP-specific genes.

| Description | pvalue | qvalue |
|---|---|---|
| cellular response to oxygen levels | 4.62E−12 | 5.60E−09 |
| cellular response to decreased oxygen levels | 4.85E−11 | 2.14E−08 |
| cellular response to hypoxia | 2.29E−10 | 7.62E−08 |
| leukocyte migration involved in inflammatory response | 5.16E−07 | 2.69E−05 |
| regulation of epithelial cell apoptotic process | 9.11E−07 | 4.29E−05 |
| response to triacyl bacterial lipopeptide | 0.003529 | 0.01902 |

Notes:
Statistical Parameters:
Description: names of the function performed by Gdf15 + MP-specific genes
pvalue/qvalue: Statistical significance of pathway enrichment
Significance threshold: $p < 0.01$, $q < 0.05$

visceral infection and inflammation.

### 3.2. DNN-SHAP framework revealed impaired aerobic respiration in scFascia MPs and identified potential crosstalk with CPTCs in dysentery

To identify the most direct transcriptional alterations in scFascia cells under Salmonella-induced dysentery, we employed annotations from the Broad Categories for deep neural network (DNN) training and feature extraction (Table 1). Expression matrices were generated for each cell type and annotated with labels corresponding to their respective treatment groups. These matrices were subsequently used as input for a neural network–based deep learning framework (Fig. 2).We identified the optimal model through iterative training and employed DeepSHAP [24] to extract interpretable features (Fig. 3A, B, C), including key signature genes, to elucidate the underlying transcriptional dynamics.

Comparison of these features with GSEA enrichment derived from differentially expressed genes (DEGs) revealed notable discrepancies in functional predictions (Fig. 3D). Extraction of the top three GSEA terms with highest activity for each cell type under dysentery conditions showed that myeloid cells within scFascia upregulated hypoxia-responsive genes (Fig. 3F), while CPTCs displayed enrichment in pathways regulating myeloid cell differentiation (Fig. 3F).

Transmission electron microscopy imaging of scFascia was performed to assess potential functional interactions between these cell types. The results demonstrated that macrophages and CPTCs frequently co-localized in the perivascular region, forming composite structures alongside pericytes and fibroblasts (Fig. 3G). Notably, morphological evidence supported functional interactions between macrophages and CPTCs (Fig. 3G). These findings suggest that regulatory signaling between CPTCs and myeloid cells may contribute to altered respiratory activity in myeloid cells under dysentery conditions.

### 3.3. MPs of scFascia in the DG exhibit augmented proliferative potential and transcriptional changes consistent with Warburg-like metabolic reprogramming

To assess how aerobic respiration-related functions in myeloid cells influence metabolic pathways, we isolated myeloid cells from the scFascia single-cell dataset and re-annotated them using canonical immune markers specific to MPs and treatment groups (Fig. 4B, D). Unsupervised clustering partitioned MPs into four subpopulations (Fig. 4A, C), followed by single-cell metabolic pathway analysis using the scMetabolic tool [25]. This analysis revealed that MPs, constituting the majority of scFascia myeloid cells, displayed significant downregulation of the tricarboxylic acid (TCA) cycle and upregulation of gluconeogenesis pathways under dysentery compared with the healthy group (Fig. 5A). These metabolic alterations reflect the Warburg-like phenotype described in enteric bacterial infections [12], typically indicative of adaptive adjustments to enhanced cellular proliferation [15,26].

Cell cycle distribution analysis showed a marked reduction in the proportion of MPs in the G1 phase and a corresponding increase in the S phase under dysentery compared to healthy conditions (Fig. 5E). These findings indicated elevated expression of genes linked to DNA replication and proliferative capacity in scFascia MPs of the dysentery group.

### 3.4. Cell state alterations of conventional M2 subsets in macrophages associated with visceral inflammation

To further dissect the functional changes of scFascia macrophages in dysentery, we examined macrophage subpopulations. A subset of Gdf15 + macrophages co-expressing Cd163, a canonical M2 marker, was identified in the dysentery group. The UMAP distance between $Gdf15^+$ macrophages in the DG and Cd163 + cells in the HG exceeded that between $Gdf15^+$ cells and other macrophage subsets, indicating distinct phenotypic divergence. Moreover, the proportion of Gdf15 + macrophages was significantly higher in the DG than in the HG (Fig. 4C), suggesting that they represent a disease-specific macrophage state.

Cytotrace analysis of differentiation states demonstrated that Gdf15 + macrophages in the DG had the highest differentiation scores (Fig. 5B), indicating greater functional specialization than their HG counterparts. Fxyd5 and S100a11 correlated strongly with Cytotrace order (Fig. 5D). Both genes are potential downstream responders to Fgf/Wnt signal [27–29]. Fxyd5 has been shown to regulate proliferation and differentiation through the Akt/mTOR pathway, which cross-regulates with Wnt/Fgf signaling [28]. Analysis of highly expressed genes in this subset revealed activation of pathways associated with fibroblast growth factor stimuli and Wnt-mediated signaling (Fig. 4E). Genes linked to hypoxia stress response, proliferation, and inflammation were markedly upregulated (Fig. 4F; Supplementary Table 1).

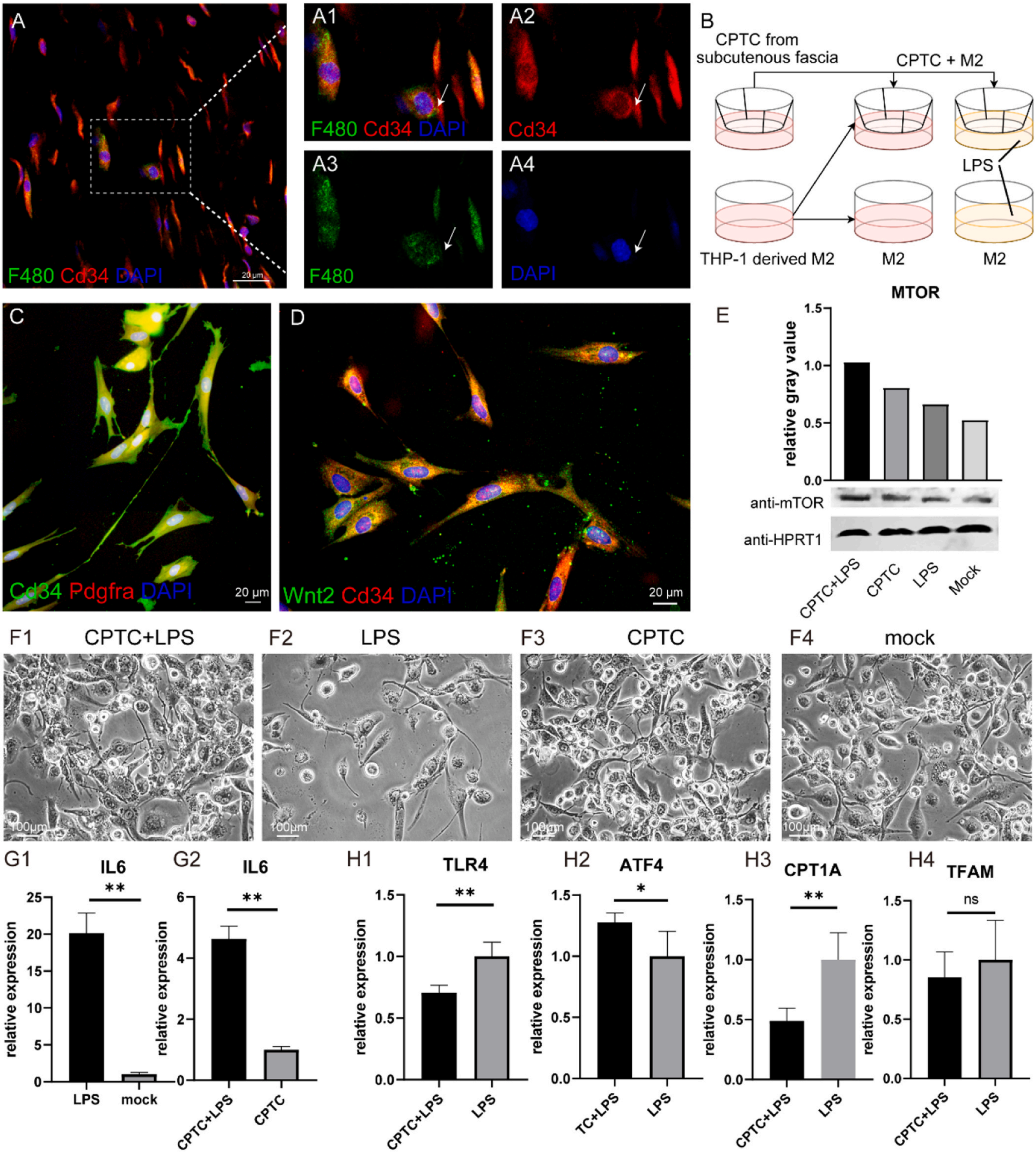

(caption on next page)

**Fig. 7.** CPTCs-M2 macrophages co-culture experiments and functional validation results.A. *In vivo* spatial distribution of macrophage populations (MPs) and CPTCs demonstrated through immunofluorescence staining of scFascia. F4/80 + macrophages (green) and CD34 + CPTCs (red) show close spatial relationships with visible direct contacts (white arrows), supporting their potential for paracrine interactions. DAPI (blue) marks nuclei. Scale bar: 20 μm.B. Experimental design and workflow of transwell co-culture system. Primary-isolated and purified CPTCs from scFascia were co-cultured with THP-1-derived M2 macrophages using transwell inserts to enable paracrine communication while preventing direct cell contact. Treatment groups included: CPTC+LPS, CPTC alone, LPS alone, and mock control.C. *In vitro* immunofluorescence identification of primary CPTCs showing uniform CD34 + (green) and PDGFRα+ (red) double-positive staining with characteristic co-localization (yellow). CPTCs display their signature elongated cellular processes (telopodes), confirming successful isolation and phenotypic retention. Scale bar: 20 μm.D. *In vitro* WNT2 secretion analysis of primary CPTCs. Extracellular WNT2 protein (green) distribution patterns emanate from CD34 + CPTCs (red) toward surrounding areas, demonstrating active paracrine factor secretion. This validates the functional capacity of isolated CPTCs for intercellular signaling. Scale bar: 20 μm.E. Western blot analysis of MTOR protein expression in M2 macrophages across treatment groups. Relative gray values normalized to HPRT1 internal reference showed the pattern: CPTC+LPS > CPTC > LPS > mock control, indicating CPTCs-mediated enhancement of MTOR signaling pathway activation.F1-F4. Morphological assessment of M2 macrophages under different treatment conditions (scale bar: 100 μm):**F1 (CPTC+LPS)**: Intermediate morphology between LPS-only and control groups, with moderated activation features**F2 (LPS alone)**: Significant morphological changes including amoeboid shape, increased lamellipodia, cytoplasmic spreading, and vacuolization, characteristic of pro-inflammatory activation**F3 (CPTC alone) & F4 (Mock)**: Elongated/spindle-shaped morphology typical of resting anti-inflammatory M2 phenotype, with uniform cell size and clear boundariesG1-G2. IL-6 mRNA expression analysis by qPCR:**G1**: LPS stimulation alone induced 20-fold upregulation of IL-6 expression (** $p < 0.01$)**G2**: CPTCs co-culture significantly attenuated LPS-induced IL-6 upregulation to 4.63-fold (** $p < 0.01$), demonstrating anti-inflammatory modulationH1-H4. Gene expression analysis of key metabolic and signaling markers:**H1**: CPTCs co-culture significantly suppressed TLR4 mRNA expression (** $p < 0.01$), indicating reduced LPS receptor availability**H2**: CPTCs co-culture significantly upregulated ATF4 mRNA expression (* $p < 0.05$), suggesting enhanced oxidative stress response**H3**: CPTCs co-culture significantly inhibited CPT1A mRNA expression (** $p < 0.01$), indicating altered fatty acid oxidation**H4**: TFAM mRNA expression showed no significant changes (ns), suggesting preserved mitochondrial biogenesis capacity.

### *3.5. Intercellular communication patterns within scFascia under dysentery conditions reveal enhanced macrophage activity signatures associated with CPTCs signaling networks*

CellCall [30] analysis was performed to interrogate intercellular communication in scFascia. CPTCs exhibited robust signaling activity toward macrophages (Fig. 6A), with overall signaling intensity significantly higher in the DG than in the HG. Prominent ligand–receptor interactions included Wnt2/Wnt5a–Lrp6, Fgf10–Fgfr1, and Rspo3–Znrf3 (Fig. 6B–D). In addition, fibroblasts in the DG uniquely communicated with macrophages via the Rspo1–Znrf3 axis (Fig. 6B–D). These interactions predominantly involved Wnt/Fgf signaling, which is known to promote proliferation.

Dissection of macrophage subpopulation communication patterns identified scFascia Gdf15⁺ macrophages as the main recipients of Wnt/Fgf signals in the DG (Fig. 6F). These macrophages displayed gene expression profiles overlapping with pathways associated with host responses to bacterial infection (Table 2) [31]. It is hypothesized that the distal metabolic alterations observed arise from secondary signaling cascades rather than direct Salmonella association.

We employed PySCENIC [32] to infer transcription factor (TF) activity from the scFascia single-cell expression matrix to explore the underlying regulatory drivers. Analysis of MP TF activity revealed significant upregulation of three TFs downstream of Wnt/Fgf signaling in the DG: Fosl1, Nfkb2, and Atf4 (Fig. 6E; Supplementary Figure 2 A). Functional relevance was validated by comparing target gene expression between groups. Differential expression and GSEA indicated that target genes were enriched for proliferation and DNA replication functions in the DG, while genes linked to aerobic respiration were downregulated (Supplementary Figure 2B, C). These results confirmed that CPTC-derived signals induced proliferative transcriptional programs in MPs of scFascia under dysentery conditions.

To validate robustness, cross-validation experiments were conducted on integrated single-cell data across biological replicates. Neural network cross-validation (Supplementary Figure 1C–F, 2D–F) showed that both G1 ($DG_1/HG_1$) and G2 ($DG_2/HG_2$) models achieved discrimination accuracies of 96.44 % and 98.47 %, respectively, in double-blind testing, with F1-scores between 0.97 and 0.99 across cell types (Supplementary Table 1). These results demonstrate that disease-associated scFascia expression patterns are highly reproducible across samples and can serve as reliable biomarkers for visceral inflammation, as captured by machine learning models.

### *3.6. CPTCs enhanced macrophage anti-inflammatory capacity by suppressing LPS sensitivity at the transcriptional level, while promoting MTOR expression at the post-transcriptional level*

Immunofluorescence staining confirmed the close spatial association between MPs and CPTCs in scFascia *in vivo* (Fig. 7A). In primary cultures of CPTCs isolated from scFascia, these cells retained their Cd34⁺/Pdgfra⁺ phenotype and Wnt2 secretion capacity (Fig. 7C–D), enabling the establishment of an *in vitro* model mimicking the *in vivo* CPTC–MP microenvironment. Transwell co-culture experiments were conducted using primary CPTCs and THP-1-derived M2 macrophages (Fig. 7B). This system simulated the CPTC–MP immune microenvironment under exposure to high-concentration LPS (1 μg/mL) for 24 h to induce pro-inflammatory reprogramming of macrophages.

CPTC co-culture significantly attenuated LPS-induced macrophage inflammatory responses, as shown by the preservation of anti-inflammatory morphology (Fig. 7F) and markedly reduced IL-6 mRNA upregulation (20-fold vs. 4.63-fold, $p < 0.01$) (Fig. 7G). Western blot analysis further demonstrated enhanced MTOR expression in the CPTC+LPS group compared to LPS treatment alone (Fig. 7E).

At the mechanistic level, CPTCs suppressed macrophage pro-inflammatory programming by downregulating TLR4 and CPT1A expression ($p < 0.01$), while upregulating ATF4 ($p < 0.05$), without altering TFAM levels (Fig. 7H). These results indicate that CPTCs remodel the immune microenvironment by transcriptionally modulating macrophage inflammatory responses and enhancing anti-inflammatory capacity through post-transcriptional promotion of MTOR expression.

## 4. Discussion

In this study, Salmonella dysentery induced shifts in the expression profiles of scFascia cell types, demonstrating that visceral inflammation can significantly influence distal cell states. Within our deep-learning framework, each scFascia cell type contributed distinct functional roles. These insights were enabled by the ability of deep learning to uncover nonlinear relationships between features and labels [33]. The single-cell data encompassed complex biological processes characterized by pronounced nonlinearity that typically require advanced analytical approaches for identification [34].By mining nonlinear features in high-dimensional data, and through single-cell metabolic analysis, we identified transcriptional changes consistent with Warburg-like reprogramming in scFascia macrophages triggered by visceral inflammation—changes that were obscured by the noise of linear features in conventional analyses (Fig. 3D vs. Fig. 3E).

Although our single-cell analysis used a limited number of biological replicates (two per group), this exploratory design aligns with hypothesis-generating investigations. We acknowledge that the small sample size may limit statistical power and the generalizability of our deep-learning models; therefore, the findings should be viewed as preliminary and in need of validation in larger cohorts. To support our computational predictions, additional validation experiments were performed as detailed below.

Subsequent single-cell functional analyses and co-culture experiments indicated that CPTCs substantially modulate macrophage functional states via autocrine/paracrine mechanisms, promoting a shift of macrophage populations toward an anti-inflammatory phenotype:

1. CPTCs and macrophages exhibited close interactions *in vivo* and *in vitro*. Cell–cell communication and transcription-factor inference analyses revealed a complex CPTC–MP network closely associated with the regulation of Warburg-like metabolic reprogramming in scFascia macrophages during visceral inflammation.
2. CPTCs mitigated LPS-induced morphological alterations in macrophages by modulating the mTOR-associated regulatory network through autocrine and paracrine mechanisms, while simultaneously enhancing overall mTOR expression. By downregulating TLR4, they reduced macrophage sensitivity to LPS, thereby strengthening the anti-inflammatory capacity of macrophage populations [35].
3. CPTCs partially enhanced glycolytic function in M2 macrophages through CPT1A downregulation [36], and modestly upregulated the stress-response gene ATF4 [37], consistent with glycolysis-driven accumulation of reactive oxygen species (ROS) in macrophage clusters [38]. Notably, unchanged expression of mitochondrial transcription factor A (TFAM) supported alignment with classical LPS-induced polarization paradigms, indicating no large-scale initiation of mitochondrial biogenesis [39].

Prior studies identify the MTOR pathway as decisive for macrophage polarization, particularly in promoting M2-type phenotypes [35], whereas MTOR inhibition generally drives pro-inflammatory phenotypes [40]. Our findings show that CPTCs markedly increase MTOR expression in co-cultured macrophages, implicating MTOR-mediated metabolic reprogramming in the modulation of immune functions [41]. Under LPS stimulation, MTOR upregulation was further amplified in CPTC-containing microenvironments, highlighting a central role in regulating a Warburg-like metabolic phenotype (enhanced glycolysis under aerobic conditions) [42] In parallel, characteristic Wnt secretion by CPTCs is known to suppress the NF-κB pathway, providing a fundamental anti-inflammatory control [43]. Collectively, CPTCs likely potentiate basal glycolytic activity in macrophages through MTOR upregulation—rather than via direct pro-inflammatory signaling—thereby fostering metabolic plasticity and functional extension in M2 macrophages. This interpretation is consistent with documented roles of MTOR/Akt in metabolic reprogramming during M2 polarization and tissue repair [44]. Based on these findings, the potential impact of CPTCs should be considered in both tumor-associated macrophage (TAM)–driven inflammatory processes and anti-inflammatory mechanisms in skin wound healing [45]. Moreover, therapeutic strategies targeting inflammatory regulatory pathways, such as canonical Wnt signaling, should be evaluated to facilitate disease recovery [46].

A further issue highlighted here is the objective impact of visceral inflammation on distal tissues. The sole intervention was induction of dysentery and acute inflammation via high-dose oral Salmonella. Subcutaneous fascial tissue exhibited significant biological changes attributable to systemic manifestations of acute inflammation. Comparable body surface–visceral correlations have been reported in other clinical scenarios [47]. Thus, scFascia may serve as a potential biomarker reflecting the progression of systemic inflammation. MPs and CPTCs, as predominant cytological components, were identified within a fascia–visceral inflammatory crosstalk pattern (FVICP) under dysbiosis. In this study, the FVICP is characterized by the following features:

1. Systemic dissemination of visceral pathological signals: the FVICP is driven by the spread of signals from visceral organs and by extensive interactions between the circulatory system and microenvironments of remote responding tissues.
2. Synchronization and mapping of immune states: immune states in certain visceral cells are mirrored in distal tissues; for example, transcriptional changes consistent with Warburg-like reprogramming were observed in scFascia macrophages during intestinal dysentery [13].
3. Pathological homeostasis in distal microenvironments: remote immune microenvironments establish a corresponding pathological equilibrium in response to visceral dysregulation.

Based on these FVICP characteristics, we hypothesize that cellular components in superficial tissues have potential diagnostic value. On the one hand, systemic inflammation is common, and similar inflammatory milieus are likely to elicit convergent fascial cellular responses. On the other hand, in cross-validation analyses of fascial cells, global fascial changes were reliably recognized by DNN models, yielding promising predictive performance. This logic parallels approaches that train machine-learning models on tumor single-cell data using well-annotated features and then leverage the generalization capacity of salient features to classify previously unseen data [48].

While our findings support the potential of fascia-derived single-cell transcriptomics for monitoring inflammatory states, several limitations should be addressed in future work:

1. Clinical translation challenges: single-cell sequencing faces barriers in accessibility, cost, and standardization for routine diagnostics; interindividual variability in fascia composition will necessitate reference ranges and workflow adaptations.
2. Evidence for metabolic reprogramming: conclusions regarding distal Warburg-like reprogramming are transcriptomic and would benefit from direct metabolic-flux validation.
3. Scope of the FVICP concept: the FVICP was derived from a single bacterial infection model and requires validation across diverse inflammatory contexts (viral infections, sterile inflammation, human disease) before broad generalization.4. *In vitro* validation constraints: THP-1–derived macrophages were used rather than primary rat macrophages, and the proposed signaling mechanisms require direct pathway validation beyond transcriptional correlations.

## Authorship Change Statement

This revision includes the addition of Yeping Shi as the second author. Ms. Shi has made substantial contributions to the investigation, formal analysis, and validation in the CPTC-MP co-culture assay that were critical for addressing reviewer comments. All authors have approved this change as documented in the accompanying Authorship Change Consent Form.

## Declaration of Responsibility

I, Qiusheng Chen, as the corresponding author, hereby state that I have collected all necessary consent forms from each co-author and confirm that the author contributions listed above accurately reflect each author's involvement in this manuscript. I take responsibility for the integrity of this author statement and verify that all co-authors have reviewed and approved this document.

## CRediT authorship contribution statement

**Qiusheng Chen:** Writing – review & editing, Writing – original draft, Supervision, Project administration, Funding acquisition, Data curation, Conceptualization. **Yeping Shi:** Validation, Investigation, Formal analysis. **Haixiang Huang:** Writing – review & editing, Writing – original draft, Visualization, Validation, Methodology, Investigation, Formal analysis, Data curation, Conceptualization. **Lu Mei:** Validation, Investigation. **Jianming Yue:** Validation, Investigation. **Zhenwei Zhang:** Validation, Investigation. **Bingbing Shen:** Validation, Investigation.

## Declaration of Competing Interest

The authors declare that they have no known competing financial interests or personal relationships that could influence the work reported in this paper.

## Acknowledgments

We extend our sincere gratitude to Professor Xu-Dong Zhu of Nanjing Agricultural University for his generous assistance and invaluable contributions throughout the course of this study. We also acknowledge the financial support provided by the National Natural Science Foundation of China (NSFC) under grant number 31872433, which made this research possible.

## Appendix A. Supporting information

Supplementary data associated with this article can be found in the online version at doi:10.1016/j.csbj.2025.09.040.

## Data Availability

All single-cell sequencing data generated during this study are available through the Gene Expression Omnibus (GEO) repository under accession number GSE261489, and the China National Center for Bioinformatics repository under accession number CRA020447. Additional data generated during experimental validation are included in the manuscript and supplementary materials. Raw data from the co-culture experiments are available from the corresponding author upon reasonable request. The scripts utilized within the single-cell sequencing pipeline and model training can be accessed from Github repo: https://github.com/NJAU-vet-QSChen-s-lab/FVICP.